\RequirePackage{lineno}
\documentclass[aps,prl,reprint,showpacs,nofootinbib]{revtex4-1}

\usepackage{times}
\usepackage{amsmath}
\usepackage{graphicx}
\usepackage{epsfig}
\usepackage{dcolumn}
\usepackage{bm,color}
\usepackage{tabularx}
\usepackage{multirow}
\usepackage{longtable}

\newcommand{\BR}{{\cal B}}
\newcommand{\eff}{\varepsilon}
\newcommand{\EE}{e^+e^-}

\newcommand{\ks}{K_{S}^{0}}
\newcommand{\ppp}{\pi^+\pi^-\pi^0}

\newcommand{\pimp}{\pi^\mp}
\newcommand{\rhopm}{\rho^\pm}
\newcommand{\etac}{\eta_c}
\newcommand{\jpsi}{J/\psi}
\newcommand{\hc}{h_c}
\newcommand{\zc}{Z_c(3900)}
\newcommand{\zcpm}{Z_c(3900)^\pm}
\newcommand{\zcp}{Z_c(4020)}
\newcommand{\zcppm}{Z_c(4020)^\pm}

\def\Journal#1#2#3#4{{#1} {\bf #2}, #3 (#4)}

\def\EPJC{Eur. Phys. J. C}

\begin{document}

\title{
\boldmath Study of $e^+e^-\to\pi^+\pi^-\pi^0 \eta_c$ and evidence for $\zcpm$ decaying into $\rho^{\pm}\etac$}

\author{
M.~Ablikim$^{1}$, M.~N.~Achasov$^{10,d}$, S. ~Ahmed$^{15}$, M.~Albrecht$^{4}$, M.~Alekseev$^{55A,55C}$, A.~Amoroso$^{55A,55C}$, F.~F.~An$^{1}$, Q.~An$^{52,42}$, Y.~Bai$^{41}$, O.~Bakina$^{27}$
% ...
\\
\vspace{0.2cm}
(BESIII Collaboration)\\
\vspace{0.2cm} {\it
$^{1}$ Institute of High Energy Physics, Beijing 100049, People's Republic of China\\
% ...
$^{4}$ Bochum Ruhr-University, D-44780 Bochum, Germany\\
% ...
$^{10}$ G.I. Budker Institute of Nuclear Physics SB RAS (BINP), Novosibirsk 630090, Russia\\
% ...
$^{15}$ Helmholtz Institute Mainz, Johann-Joachim-Becher-Weg 45, D-55099 Mainz, Germany\\
% ...
$^{27}$ Joint Institute for Nuclear Research, 141980 Dubna, Moscow region, Russia\\
% ...
$^{41}$ Southeast University, Nanjing 211100, People's Republic of China\\
$^{42}$ State Key Laboratory of Particle Detection and Electronics, Beijing 100049, Hefei 230026, People's Republic of China\\
% ...
$^{52}$ University of Science and Technology of China, Hefei 230026, People's Republic of China\\
% ...
$^{55}$ (A)University of Turin, I-10125, Turin, Italy; (B)University of Eastern Piedmont, I-15121, Alessandria, Italy; (C)INFN, I-10125, Turin, Italy\\
% ...
$^{d}$ Also at the Novosibirsk State University, Novosibirsk, 630090, Russia\\
% ...
}
}

\date{\today}

\begin{abstract}

We study the reaction $e^{+}e^{-} \to \pi^{+}\pi^{-}\pi^{0}\eta_{c}$ for the first time using data samples collected with the
BESIII detector at center-of-mass energies $\sqrt{s}=4.226$, $4.258$, $4.358$, $4.416$, and $4.600$~GeV. Evidence of this process 
is found and the Born cross section $\sigma^{B}(\EE\to \pi^{+}\pi^{-}\pi^{0}\eta_{c})$, excluding $\EE\to\omega \eta_{c}$ and $\eta \eta_{c}$, is measured  to be $(46^{+12}_{-11} \pm 10)\rm \,pb$ at $\sqrt{s}=4.226$\,GeV.
Evidence for the decay $\zcpm\to\rhopm\etac$ is reported at $\sqrt{s} = 4.226$~GeV with a significance of $3.9\sigma$, including  systematic uncertainties, and the Born cross section times branching fraction 
$\sigma^{B}(\EE\to \pimp\zcpm)\times \BR(\zcpm\to\rhopm\etac)$ is measured to be $(48 \pm 11 \pm 11)\,\rm{pb}$, which indicates that 
$\EE\to\pimp\zcpm\to\pimp\rhopm\etac$ dominates the $\EE\to\pi^{+}\pi^{-}\pi^{0}\eta_{c}$ process. The $\zcpm\to\rhopm\etac$ signal 
is not significant at the other center-of-mass energies and the corresponding upper limits are determined. In addition, no significant signal 
is observed in a search for $\zcppm\to \rho^{\pm}\etac$ with the same data samples. The ratios $R_{\zc}=\BR(\zcpm\to \rho^{\pm} \etac)/\BR(\zcpm\to \pi^{\pm}
\jpsi)$ and $R_{\zcp}=\BR(\zcppm\to \rho^{\pm}\etac)/\BR(\zcppm\to \pi^{\pm} \hc)$ are obtained and  compared with different theoretical interpretations of the $\zcpm$ and $\zcppm$.

\end{abstract}


\maketitle


The charged charmonium-like states
$\zcpm$~\cite{Ablikim:2013mio,Liu:2013dau,Ablikim:2013xfr} and
$\zcppm$~\cite{Ablikim:2013wzq,Ablikim:2013emm} were first observed
in 2013. Although their observed properties indicate they are
not conventional mesons consisting of a quark-antiquark pair,
their exact quark configurations are still unknown. Several
models have been developed to describe their inner
structure~\cite{reviews}, including loosely bound hadronic
molecules of two charmed mesons~\cite{Voloshin:1976ap}, compact
tetraquarks~\cite{Maiani:2004vq,Wang:2013vex}, and
hadro-quarkonium~\cite{Voloshin:2007dx,Dubynskiy:2008mq}.

It has recently been suggested that the relative decay rate of $Z_{c}$ states, 
such as $Z_{c}(3900)\to \rho\etac$ to $\pi\jpsi$ (or $Z_{c}(4020)\to
\rho\etac$ to $\pi\hc$), can be used to discriminate between the tetraquark
and meson molecule scenarios~\cite{Esposito:2014hsa}. 
In Ref.~\cite{Esposito:2014hsa},
the predicted ratio $R_{Z_c(3900)}=\BR(\zc\to \rho\eta_c)/\BR(\zc\to \pi\jpsi)$ 
is $230_{-140}^{+330}$ or $0.27_{-0.17}^{+0.40}$ based on the diquark-antidiquark tetraquark model, 
depending on how the spin-spin interaction outside the
diquarks is treated.  On the other hand, using Non-Relativistic Effective Field Theory techniques, this ratio is only $0.046_{-0.017}^{+0.025}$ 
 if we assume the $Z_c(3900)$ is a meson molecule state. 
Similarly, the predicted ratio of $R_{Z_c(4020)}=\BR(\zcp\to \rho\etac)/\BR(\zcp\to \pi\hc)$ is
$6.6_{-5.8}^{+56.8}$ in the tetraquark model, but only $0.010_{-0.004}^{+0.006}$ 
in the meson molecule model~\cite{Esposito:2014hsa}.   
However, the well-separated predictions for $R_{Z(3900)}$ and $R_{Z(4020)}$,  shown above,
could move closer or even overlap according to different theoretical approaches.
Within QCD sum rule approaches~\cite{Faccini:2013lda, Agaev:2016dev, Dias:2013xfa,
Wang:2017lot} and convaraint quark model approaches~\cite{Goerke:2016hxf} to the tetraquark scenario, the predicted value of $R_{Z_c(3900)}$ can vary from 0.66 to 1.86.
Furthermore, different approaches to the meson molecule model~\cite{Patel:2014zja, Ke:2013gia, Goerke:2016hxf} can lead to predictions for $R_{Z_c(3900)}$ from $6.8\times 10^{-3}$ to $1.8$.
Consequently, the capability to separate the molecular and tetraquark models is currently model-dependent. 
In the hadron-charmonium model, the $Z_{c}(3900)$ is treated as a $J/\psi$ embedded in an
S-wave spinless excitation of light-quark matter and consequently the transition $Z_{c}(3900)\to \rho\etac$
is expected to be suppressed compared to $Z_{c}(3900)\to\pi\jpsi$.
A search for the decays
of $Z_{c}(3900)$ or $Z_{c}(4020)$ to $\rho\etac$ thus offers 
an important opportunity to 
discrimate among the wide range of theoretical predictions.


In this Letter, we first report a search for the process $\EE\to\pi^{+}\pi^{-}\pi^{0}\eta_{c}$.  
Then, based on the first step, we study the subprocesses
$\EE\to\pi Z_c(3900)^{\pm}$; $Z_{c}(3900)^{\pm}\to\rhopm\etac$ and  
$\EE\to\pi Z_c(4020)^{\pm}$; $Z_{c}(4020)^{\pm}\to\rhopm\etac$.
 We use data samples collected with the BESIII detector~\cite{bes3-detector} at
center-of-mass (c.m.) energies above 4~GeV, as listed in
Table~\ref{Born_CS_pirho}. 
The c.m.~energies are measured
using the $e^{+}e^{-}\to\mu^{+}\mu^{-}$ process with an uncertainty of $\pm
0.8$~MeV~\cite{Ablikim:2015zaa}. The beam spread is measured to be 1.6 MeV. 

The design and performance of the
BESIII detector are given in Ref~\cite{bes3-detector}.
A {\sc geant4}-based~\cite{geant4} Monte Carlo (MC) simulation
software package
is used to optimize
event selection criteria, determine the detection efficiencies,
and estimate the backgrounds. 
At each energy, the signal events
are generated according to phase space using {\sc
evtgen}~\cite{evtgen}. Initial state radiation (ISR) is
simulated with {\sc kkmc}~\cite{kkmc}, and final state radiation is handled with {\sc
photos}~\cite{photos}. 


Charged tracks, photons and $\ks$ candidates are reconstructed using
the standard criteria of the BESIII experiment~\cite{Ablikim:2017pg}.
Candidate $\pi^0$ and $\eta$ decays to $\gamma \gamma$ are reconstructed from pairs of photons with 
invariant mass in the range [0.120,~0.145]~GeV/$c^{2}$ for
the $\pi^{0}$ and [0.50,~0.57]~GeV/$c^2$ for the $\eta$.
To improve the resolution, a
one-constraint (1C) kinematic fit is imposed on the selected
photon pairs to constrain their invariant mass to the nominal
$\pi^{0}$ or $\eta$ mass~\cite{Tanabashi:2018oca}.

The $\eta_{c}$ candidates are reconstructed using nine hadronic
decays: $p\bar{p}$, $2(K^+K^-)$, $K^+ K^- \pi^+ \pi^-$,  $K^+ K^- \pi^0$, $p \bar{p} \pi^0$,
$\ks K^\pm \pi^\mp$,  $\pi^+ \pi^- \eta$, $K^+ K^- \eta$, and $\pi^+ \pi^- \pi^0 \pi^0$. 
 All combinations with invariant mass in the range [2.7,~3.2]~GeV/$c^2$ are kept within each event.
The signal region for the $\eta_c$ candidates is defined as [2.95,~3.02]~GeV/$c^2$ and
the sidebands as [2.78,~2.92] and [3.05,~3.19]~GeV/$c^2$.

After the above selection, a four-constraint (4C)
kinematic fit is performed for each event, and the $\chi^{2}$ of
the fit ($\chi^2_{\rm 4C}$) is required to be less than 40 to
suppress backgrounds. In each event, the mass of each track
(excluding $K_S^0$ daughters) is taken to be that of the kaon, 
pion or proton, depending on the decay mode under study. Finally, 
only the combination of mass assignments with the minimum $\chi^2_{min} \equiv
\chi^2_{\rm 4C}+\chi^2_{\rm 1C}+\chi^2_{\rm PID}+\chi^2_{\rm
vertex}$  is kept. Here, $\chi^2_{\rm 1C}$ is
the $\chi^2$ of the 1C fit for $\pi^0$ ($\eta$), $\chi^2_{\rm
PID}$ is the sum of the $\chi^{2}$ for the PID
of all charged tracks, and $\chi^2_{\rm vertex}$ is the $\chi^2$
of the $K_S^0$ secondary vertex fit.

Inclusive MC samples with the same statistics as the data are
studied to understand the backgrounds. The major backgrounds to
$\EE\to \pi^+\pi^-\pi^{0}\eta_{c}$ are classified into two
categories. They are events from (1) charmonium(-like) states decays~(most of which include
open-charm decays, e.g. $\psi\to D^{(*)}\bar{D}^{(*)}$); and (2) the continuum process,
$\EE\to q\bar{q}$, with $q=u$, $d$, and $s$. 

By analyzing 600,000 $e^{+}e^{-}\to \pi^{+}\pi^{-}h_{c}$ MC simulation events
with $h_c$ decaying inclusively, a small enhancement in the
$\eta_{c}$ signal region is found. Using the measured cross section given in 
Ref~\cite{Ablikim:2013wzq} and the luminosity of data, its contribution, $N_{\rm bkg}^{\rm peaking}$, 
is estimated to be $8.7\pm 2.0$ at $\sqrt{s}=4.226$~GeV. The contributions at other energies are
estimated in a similar way. 

To suppress background events with charmed mesons, events are rejected if a $D$
meson candidate is reconstructed in one of its five decay
modes: $D^{0}\to K^{\pm} \pi^{\mp}$, $D^{0}\to K^{\pm} \pi^{\mp}
\pi^0$, $D^{\pm}\to K^{\pm} \pi^{\mp} \pi^{\pm}$, $D^{\pm}\to
K_{S}^{0} \pi^{\pm}$, and $D^{\pm}\to K_{S}^{0} \pi^{\pm} \pi^0$.
To accomplish this, we require the invariant mass of $D^{0}$($D^{\pm}$) candidates
to be outside the region $m(D^{0})\pm24$\,MeV ($m(D^{\pm})\pm10$\,MeV).
To reduce the continuum background, events with a $K^*(892)\to K\pi$, an $\omega\to \ppp$, or an
$\eta\to \ppp$ candidate are removed by requiring $|M(K\pi)-m(K^{*})|>32$\,MeV, $|M(\ppp)-m(\omega)|>26$\,MeV,
and $|M(\ppp)-m(\eta)|>10$\,MeV, respectively. Here, $m(D^{0})$,
$m(D^{\pm})$, $m(K^{*})$, $m(\omega)$ and $m(\eta)$ are the nominal masses of
the corresponding states. 

The mass windows for the background veto mentioned above and the $\chi^{2}$
requirement of the 4C kinematic fit are determined by optimizing
the figure-of-merit (FOM), which is defined as ${\rm FOM} =
S/\sqrt{S+B}$. Here, $S$ is the number of signal events from the
MC simulation assuming $\sigma(\EE\to \ppp\etac)=50~{\rm pb}$,
which is evaluated from a measurement with unoptimized selection
criteria. $B$ is the number of background events obtained
from the $\eta_{c}$ sidebands in the data and extrapolated to the signal region linearly. 
The optimization is performed through iterations until all the selection criteria
become stable.

\begin{figure}[htbp]
\epsfig{file=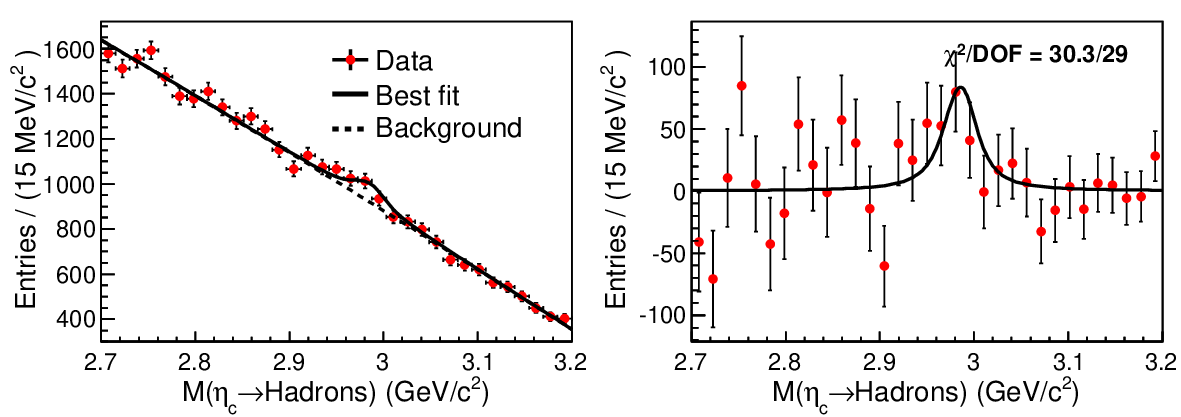,width=8.7cm}
\caption{Invariant mass distributions of the $\eta_c$ candidates
summed over nine channels in $\EE\to \ppp\etac$ at
$\sqrt{s}=4.226$~GeV (left panel), and the signal after background
subtraction (right panel). Dots with error bars are the data,
solid lines are the total fit, and the dotted line is background.
\label{fig:3pietac_simul_fit_data}}
\end{figure}

To obtain the $\ppp\etac$ yield, the invariant mass distributions
of the $\eta_{c}$ candidates in the nine decay modes are fitted
simultaneously using an unbinned maximum likelihood method. In the
fit, the $\eta_c$ signal shape is determined from MC simulation 
and is described with a constant-width Breit-Wigner function
(mass and width are fixed to the world average
values~\cite{Tanabashi:2018oca}) convolved with a Crystal Ball function, which represents instrumental
resolution. The background is described with a second order Chebyshev Polynomial (CP). 
Both the signal and background shapes are channel dependent,
but the relative signal yields among all the channels are
constrained by branching fractions and efficiencies~\cite{Ablikim:2017pg}. 
The total signal yield of the nine channels is labeled $N_{\rm obs}$, which is
shared for all the channels and required to be positive. 
The free parameters in the fit include $N_{\rm obs}$ and the
background yield and shape parameters for each decay mode. 
Figure~\ref{fig:3pietac_simul_fit_data}(left) shows the fit results at
$\sqrt{s}$ = 4.226~GeV projected onto the sum of events from all nine $\etac$ decay modes.
Figure~\ref{fig:3pietac_simul_fit_data}(right) shows the background-subtracted distribution.
The total signal yield is $333^{+83}_{-80}$ with a statistical
significance of $4.2\sigma$, which is obtained by comparing the
change of the log-likelihood value $\Delta(-\ln L) = 9.0$ 
with and without the $\ppp\eta_{c}$ signal in the fit with one degree of freedom. The same selection criteria are
applied to the other data sets, but no significant signals are observed.

The Born cross section of the $\EE\to \ppp\etac$ reaction is
calculated using
\begin{equation}\label{eq1}
\sigma^{\rm B}(\EE\to \ppp\etac)=\frac{N_{\rm sig}}
     {\mathcal{L}(1+\delta)\frac{1}{|1-\Pi|^2} \sum_{i}\eff_i \BR_i},
\end{equation}
where $N_{\rm sig}=N_{\rm obs}-N_{\rm bkg}^{\rm peaking}$ is the
number of signal events after the peaking background subtraction;
$\mathcal{L}$ is the integrated luminosity; $(1+\delta)$ is the
ISR correction factor, assuming the $\ppp\etac$ signal is from 
$Y(4260)$ decays~\cite{Tanabashi:2018oca}; and $\frac{1}{|1-\Pi|^2}$ is
the vacuum-polarization factor~\cite{Jegerlehner:1985gq}. The
cross sections and the numbers used for their calculation are listed
in Table~\ref{Born_CS_pirho} for all energy points.  The upper
limits of the cross sections at 90\% confidence level (C.L.)  
are determined using
a Bayesian method, assuming a flat prior in $\sigma^{\rm B}$. The
systematic uncertainties are incorporated into the upper limit by
smearing the probability density function of the cross
section~\cite{Ablikim:2017pg}. The corresponding results for
$\sigma^{\rm B}_{\rm U.L.}$ are also listed in
Table~\ref{Born_CS_pirho}.

\begin{table}[htbp]
\caption{\label{Born_CS_pirho} The Born cross section
($\sigma^{\rm B}$) for the $\EE\to \ppp\etac$ process and the
numbers that enter the calculation (see Eq.~(\ref{eq1})). 
Here, $\sqrt{s}$ is in GeV, $\mathcal{L}$ is in $\rm pb^{-1}$, $\sum \eff \BR$ is
in \% and $\sigma^{\rm B}$ is in pb.
}
  \begin{tabular}{ccccccc}
  \hline\hline
  $\sqrt{s}$   &    $\mathcal{L}$    &  $N_{\rm sig}$
    & $(1+\delta)$  & $\frac{1}{|1-\Pi|^2}$ & $\sum \eff \BR$ 
      & $\sigma^{\rm B}$($\sigma^{\rm B}_{\rm U.L.}$)        \\
  \hline
  4.226 & 1091.7 & $ 324^{+83}_{-80} $ &  0.74 & 1.056 &  0.82 & $ 46^{+12}_{-11}\pm 10  $   \\
  4.258 & 825.7 & $ 157^{+73}_{-68} $ &  0.76 & 1.054 &  0.80 & $ 30^{+14}_{-13}\pm 9  $ ( $< 67 $)  \\
  4.358 & 539.8 & $ 32^{+62}_{-24} $ &  1.03 & 1.051 &  0.62 & $ 9^{+17}_{-7}\pm 2  $ ( $< 41 $)  \\
  4.416 & 1073.6 & $ 19^{+82}_{-18} $ &  1.15 & 1.053 &  0.49 & $ 3^{+13}_{-3}\pm 1  $ ( $< 38 $)  \\
  4.600 & 566.9 & $ 0^{+28}_{-0} $ &  1.32 & 1.055 &  0.31 & $ 0^{+12}_{-0}\pm 13  $ ( $< 36 $)  \\
  \hline\hline
  \end{tabular}
\end{table}

\begin{figure}[htbp]
\epsfig{file=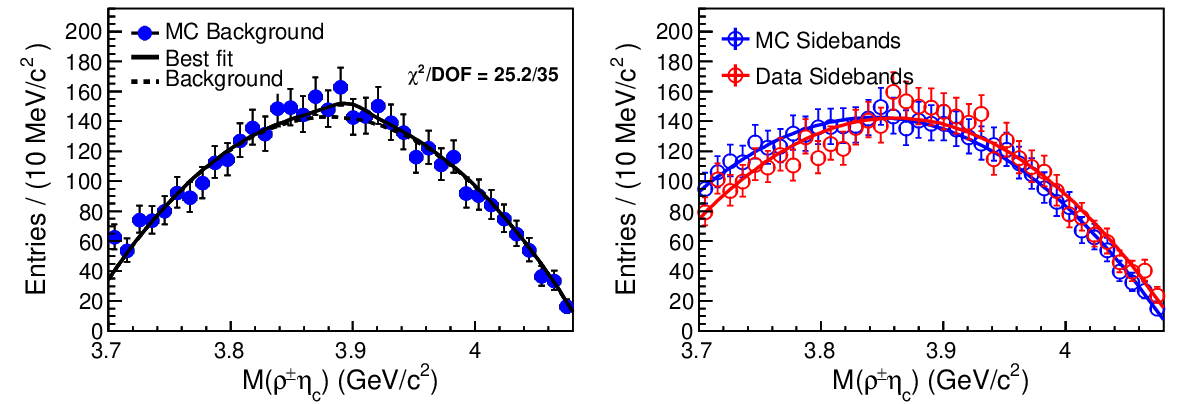,width=8.7cm}
\caption{
Left: fit to the simulated background at $\sqrt{s}=4.226$~GeV in the $\eta_c$ signal region. 
The black solid line is the best fit and dots with error bars are simulated background. Right: 
fit to the sidebands in data and MC.  The blue and red solid lines are the second order CP functions, 
the open blue and red dots with error bars are $\etac$ sidebands in MC and data (color online).  
\label{fit_validation_4230}}
\end{figure}

\begin{figure}[htbp]
\epsfig{file=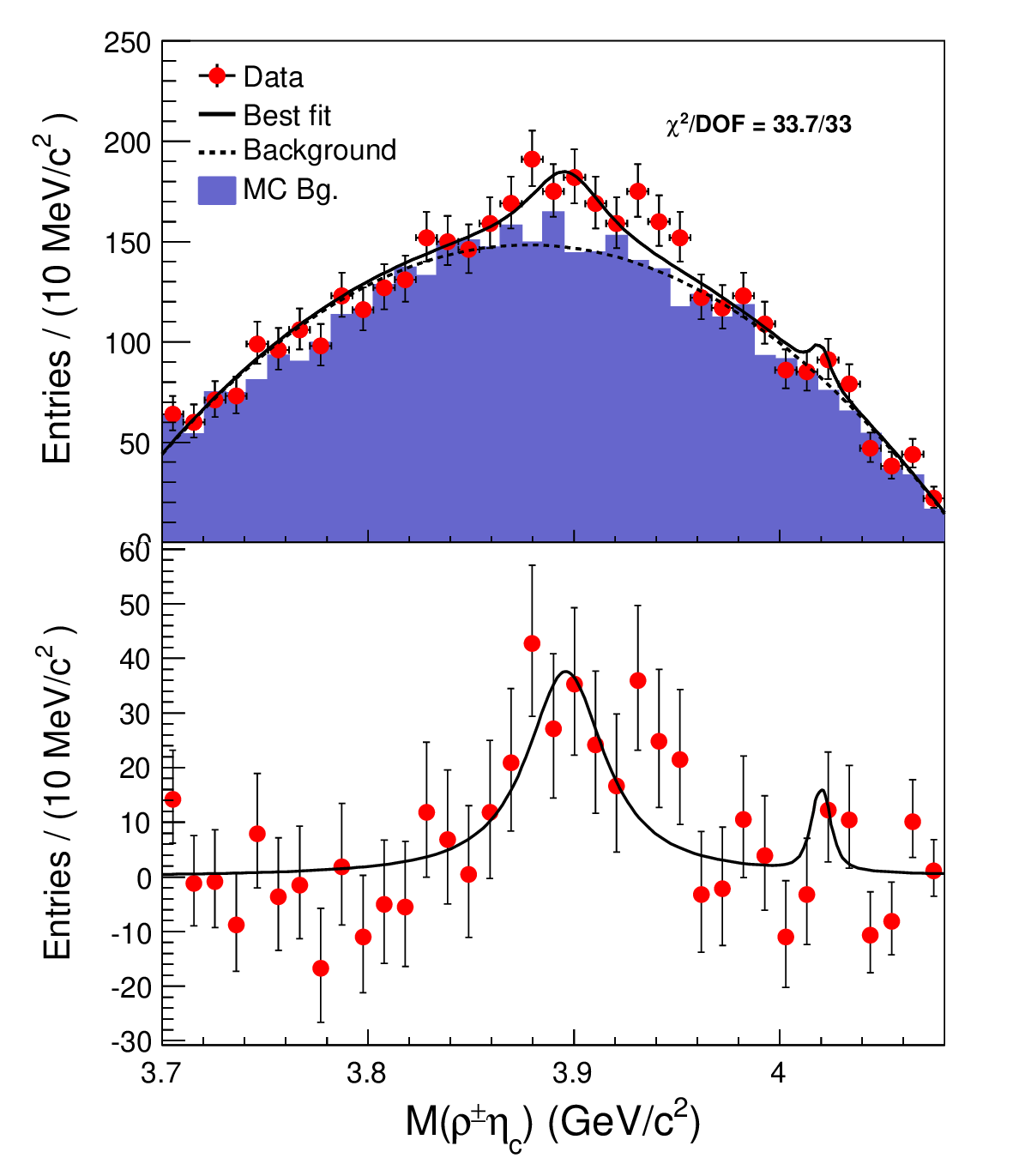,width=7cm}
\caption{The $\rho^\pm\eta_c$ invariant mass distribution summed
over nine $\eta_c$ decay channels in $\EE\to
\pi^{\mp}\rho^{\pm}\etac$ at $\sqrt{s}=4.226$~GeV. Top:  dots with error bars are
data and the shaded histogram is the simulated background. The solid line is the total fit and the dotted 
line is the background. Bottom: the same plot with the background subtracted.
\label{fit_Zc_data_4230}}
\end{figure}

The $Z_{c}(3900)^{\pm}$  and $Z_{c}(4020)^{\pm}$ signals are examined after requiring
that the invariant mass of an $\eta_{c}$ candidate is within the
$\eta_{c}$ signal region [2.95,~3.02]~GeV/$c^2$ and the invariant
mass of $\pi^{\pm}\pi^{0}$ is within the $\rho$ signal region
[0.675,~0.875]~GeV/$c^2$. Here, we don't distinguish the pions from $\eta_c$ decay or from collision and $\rho$ decay, 
therefore all possible combinations in one event  are kept to avoid bias. To suppress the combinatorial
background, the momenta of the pions from the $\rho$ decays are
required to be less than 0.8~GeV/$c$. The events in the $\etac$ sidebands and 
$\rho$ sideband, which is defined as [0.475,~0.675]~GeV/$c^{2}$, 
are investigated and no peaking structure is found. 
In addition, the simulated background events are studied~(Fig.~\ref{fit_validation_4230}~left) and show good agreement with data 
both in the $\eta_c$ signal (Fig.~\ref{fit_Zc_data_4230} top) and sideband regions (Fig.~\ref{fit_validation_4230} right). 
In the data sample, the $\zcpm$ signal is apparent, but there is no statistically significant $\zcppm$ signal.

To obtain the yields of $e^{+}e^{-}\to \pi^{\mp}Z_{c}(3900)^{\pm}\to\pi^{\mp}
\rho^{\pm}\eta_{c}$ and $e^{+}e^{-}\to \pi^{\mp}Z_{c}(4020)^{\pm}\to\pi^{\mp}
\rho^{\pm}\eta_{c}$, the invariant mass of $\rho^{\pm}\eta_{c}$ candidates in the
nine $\etac$ decay channels are fitted simultaneously using the
same method as for $e^{+}e^{-}\to \ppp\etac$. In the fit, a
possible interference between the signal and the background is
neglected. The mass and width of the $\zcpm$ are fixed to the
values from the latest measurement~\cite{Collaboration:2017njt}
and those of the $\zcppm$ are fixed to world average
values~\cite{Tanabashi:2018oca}. The mass resolution is obtained from
MC simulation and parameterized as a Crystal Ball function~\cite{Oreglia:1980cs}. 
The background is described with a second order CP function.  To validate the fit model, 
we perform a fit with the same model on the simulated background as shown in Fig.~\ref{fit_validation_4230} (left).
The signal yields of $\zcpm$ and $\zcppm$ are $48\pm46$ and $0\pm4$, respectively, and 
the statistical significance of the $\zcpm$ is $0.6\sigma$. We also fit the sideband events both 
from data and MC with the second order CP function and the function can 
describe the sidebands well as shown in Fig.~\ref{fit_validation_4230} (right).  
After the validation, we apply the fit model to data.
Figure~\ref{fit_Zc_data_4230} shows the fit to the data set taken at $\sqrt{s}=4.226$~GeV.
The total $\zcpm$ signal yield is $240^{+56}_{-54}$ events with a statistical
significance of $4.3\sigma$, and that of the $\zcppm$ is $21^{+15}_{-11}$ events
with a statistical significance of $1.0\sigma$. The signals at the
other c.m. energies are not statistically significant.

The Born cross section for $\EE\to \pimp Z_{c}^{\pm}$ with $Z_{c}^{\pm}\to \rhopm\etac$ 
is calculated using the same equation as shown in Eq.~(\ref{eq1}). The numbers used 
in the calculation and the results are listed in Table~\ref{Born_CS_pizc}.

\begin{table*}[htbp]
\caption{\label{Born_CS_pizc} Born cross sections of $\EE\to \pimp
Z_{c}(3900)^{\pm} \to \pimp\rhopm\etac$ and $\EE\to \pimp
Z_{c}(4020)^{\pm} \to \pimp\rhopm\etac$ . $\mathcal{S}$ is the
statistical significance of the signal. Other parameters are defined in the
same way as those in Table~\ref{Born_CS_pirho}. Here, $Z_c(3900)$ is labeled as $Z_c$ and $Z_c(4020)$ is labeled as $Z_c^{\prime}$.}
 {
  \begin{tabular}{ccccccccccccc}
  \hline\hline
  $\sqrt{s}$ (GeV)  &     $N_{\rm obs}^{Z_{c}}$ &  $N_{\rm obs}^{Z_{c}^{\prime}}$
    & $(1+\delta)$  & $\frac{1}{|1-\Pi|^2}$ & $\sum \eff^{Z_{c}} \BR$ (\%) & $\sum \eff^{Z_{c}^{\prime}} \BR$ (\%)
    & $\sigma^{\rm BZ_c}$ (pb)  & $\sigma^{\rm BZ_c}_{\rm U.L.}$ (pb) &  $\sigma^{\rm BZ_c^{\prime}}_{\rm U.L.}$ (pb) &   $\mathcal{S}^{Z_c}$ ($\sigma$) & $\mathcal{S}^{Z_c^{\prime}}$ ($\sigma$)\\
  \hline
  4.226 & $ 240^{+56}_{-54}$&$21^{+15}_{-11} $ &  0.74 & 1.056 &  0.59 &0.52 & $ 48^{+11}_{-11} \pm 11  $ & ... &$<14 $ & 4.3&1.0 \\
  4.258 & $ 92^{+48}_{-43}$&$0^{+11}_{-0} $ &  0.76 & 1.054 &  0.50 &0.56 & $ 28^{+15}_{-13} \pm 8  $ & $< 62$&$<6$ & 2.0&... \\
  4.358 & $ 12^{+40}_{-8}$&$0^{+15}_{-0} $ &  1.03 & 1.051 &  0.44 &0.42 & $ 5^{+16}_{-3} \pm 2  $ & $< 36$&$<14 $ & 0.3&... \\
  4.416 & $ 101^{+48}_{-44}$&$6^{+17}_{-4} $ &  1.15 & 1.053 &  0.35 &0.34 & $ 22^{+10}_{-10} \pm 5  $ & $< 44$&$<11 $ & 2.2&... \\
  4.600 & $ 0^{+11}_{-0}$&$0^{+10}_{-0} $ &  1.32 & 1.055 &  0.20 &0.21 & $ 0^{+7}_{-0} \pm 1  $ & $< 14$&$<21 $ & ... & ...\\
  \hline\hline
  \end{tabular}
  }
\end{table*}

The systematic uncertainties in the $\sigma^{\rm B}(\EE\to \ppp\etac)$
measurement originate from the uncertainty of each factor in
Eq.~(\ref{eq1}). The integrated luminosity has an uncertainty of
$1.0\%$~\cite{Ablikim:2015nan}. 
The uncertainty due to the subtraction of the $\EE\to\pi^+\pi^-\hc$ peaking background events includes both the uncertainty due to the cross section and the statistical error of the MC sample.
To estimate the uncertainty due to ISR correction, the c.m. energy dependent cross
section of $\EE\to \pi^+\pi^-J/\psi$ measured by the BESIII
experiment~\cite{Ablikim:2016qzw} is used instead of Y(4260). The 
uncertainty from the signal shape consists of the
mass resolution discrepancy between data and MC simulation and the
uncertainty of the $\eta_{c}$ resonant parameters. The former is
studied using an $\EE\to \gamma_{ISR}J/\psi$~\cite{Ablikim:2017ove}
sample and the latter is estimated by varying the $\eta_{c}$
mass and width by $\pm 1\sigma$ around the world average
values~\cite{Tanabashi:2018oca}. The uncertainty for the background
shape is estimated by changing the order of the CP function and adjusting the
 fit boundaries. The methods for estimating the uncertainties due to the
vacuum polarization and $\sum_{i}\eff_i \BR_i$ are the same as those described in
Ref.~\cite{Ablikim:2017pg}.  Furthermore,  the uncertainty due to the $\EE\to \ppp\etac$ decay
dynamics is obtained by comparing the simulations with and without the $Z_{c}$
resonance. All of the sources are assumed to be independent and added
in quadrature and the largest systematics uncertainty is that of $\sum_{i}\eff_i \BR_i$. 
The total systematic uncertainties are listed in Table~\ref{Born_CS_pirho}.

For the $\sigma^{\rm B}(\EE\to \pimp Z_{c}(3900)^{\pm}(Z_{c}(4020)^{\pm})\to
\pimp\rhopm\etac)$ measurement, the uncertainties on $\mathcal{L}$, ISR factors, $\sum_{i}\eff_i \BR_i$ 
and the vacuum polarization factor are studied following the methods described in the measurement 
of $\sigma^{\rm B}(\EE \to \pi^{+} \pi^{-} \pi^{0} \eta_{c})$. Moreover, additional  systematic
uncertainties arise from the $\rho$ and $\etac$ selections, and the
fit of the invariant mass spectrum of $\rho^{\pm}\eta_{c}$. The uncertainty
due to the $M(\pi^{\pm}\pi^{0})$ mass window is estimated by
comparing the invariant mass of $M(\omega \to \ppp)$ in data and MC assuming the mass resolution 
of $M(\ppp)$ is larger than $M(\pi^{\pm}\pi^{0})$. The discrepancy is found to be negligible. 
The uncertainty of the $\eta_{c}$ line shape is estimated by the varying the mass and width of the $\eta_c$
within the errors given by world average values~\cite{Tanabashi:2018oca}. The uncertainties affecting the fit to
the $Z_{c}(3900)^{\pm}$ ($Z_{c}(4020)^{\pm}$) are estimated with the same methods
as in the $\ppp\etac$ case. All these sources and those in the
$\sigma^{\rm B}(\EE\to \ppp\etac)$ measurement are assumed to be independent and added in quadrature. 
The uncertainties related to the fit of invariant mass of $\eta_c\to $ hadrons are excluded because they don't 
affect the  $\EE \to \pi Z_c$ measurement. The largest systematics uncertainty comes from $\sum_{i}\eff_i \BR_i$. 
The total systematic uncertainties are listed in Table~\ref{Born_CS_pizc}. 

To evaluate the effect of the systematic uncertainty
on the signal significance at $\sqrt{s}=4.226$\,GeV, we vary the
signal shape, background parametrization, and fit range, or free
the $Z_{c}$ mass, then repeat the fit. We find that the statistical 
significance of the $Z_{c}(3900)$ is always larger than $3.9~\sigma$.

In summary, using the $e^+e^-$ annihilation data at $\sqrt{s}=4.226$, $4.258$, $4.358$, $4.416$, and $4.600$~GeV,
we study the $e^+e^-\to\pi^+\pi^-\pi^0 \eta_c$ process for the first time. 
Evidence of this process is observed at $\sqrt{s}=4.226$~GeV with a significance of $4.2\sigma$ and the Born cross section $\sigma^{\rm B}(\EE\to
\ppp\etac)$ is measured to be $(46^{+12}_{-11} \pm 10)\rm \,pb$, excluding the processes $\EE\to
\omega\etac$ and $\eta\eta_{c}$.
Evidence for the $\rho^{\pm}\eta_{c}$ decay mode of the charged charmonium-like state $\zcpm$ is found
in the process $\EE\to \pimp \zcpm$ with $\zcpm\to\rhopm\etac$ from the same data set. The measured
 cross section times branching ratio $\sigma^{\rm B}(\EE\to \pimp \zcpm)\times \BR(\zcpm\to\rhopm\etac)$ is 
 $(48 \pm 11 \pm 11)\rm{\,pb}$. This result indicates that the $\EE\to \ppp\etac$ process is dominated by the subprocess $\EE\to
\pimp \zcpm\to \pimp\rhopm\etac$ (and implicitly $\EE\to\pi^0 Z_c(3900)^0 \to \pi^0 \rho^0 \etac$).
The significance of $\zcpm\to \rhopm\etac$ is $3.9\sigma$ including the
systematical uncertainty. No significant signal of $e^+e^-\to\pi^+\pi^-\pi^0 \eta_c$  is observed at
$\sqrt{s}=$4.258, 4.358, 4.416, and 4.600~GeV and 
no significant signal of  $\EE\to \pimp \zcppm$ with $\zcppm\to \rhopm\etac$ is found in any of
the data sets.  Upper limits are deterimened at 90\% C.L.

Using the results from Refs.~\cite{Ablikim:2013wzq}
and~\cite{Collaboration:2017njt}, we calculate the ratios
$R_{\zc}=\BR(\zcpm\to \rho^{\pm} \etac)/\BR(\zcpm\to \pi^{\pm}
\jpsi)$ and $R_{\zcp}=\BR(\zcppm\to \rho^{\pm}
\etac)/\BR(\zcppm\to \pi^{\pm} \hc)$. The results obtained from the measurements at
$\sqrt{s}=$4.226, 4.258, and 4.358~GeV are listed in Table~\ref{Rz},
together with the theoretical predictions for comparison.

The measured $R_{\zc}$ is closer to the calculation of the
 tetraquark model than to that of the meson molecule
model in Ref.~\cite{Esposito:2014hsa}.
The measurement is also consistent with several
other independent calculations based on the tetraquark
scenario~\cite{Faccini:2013lda, Agaev:2016dev, Dias:2013xfa,
Wang:2017lot,Goerke:2016hxf}. For the molecule model, as we mentioned before,  the
calculated $R_{\zc}$ is highly model dependent 
~\cite{Patel:2014zja, Ke:2013gia,
Goerke:2016hxf}. 
Therefore, it is necessary
to narrow down the theoretical uncertainty in the molecular
framework to have a better comparison with the measurement. In the
hadron-charmonium model, the $\BR(\zc\to \rho \etac)$ is suppressed
compared with $\BR(\zc\to \pi \jpsi)$ and therefore inconsistent with
the measurement~\cite{Voloshin:2013dpa}. Furthermore, this model
predicts a new resonance $W_{c}(3785)$, which can be produced via
$e^{+}e^{-}\to \rho W_{c}\to \rho\pi\eta_{c}$, the same final
state we analyzed here. As we found that the $\EE\to \ppp\etac$
process is saturated by $\EE\to \pi \zc\to \rho\pi\etac$, we can conclude
that the production of the $W_c$, if present, is small compared to $\EE\to \pi
\zc$.

\begin{table}[htbp]
\caption{\label{Rz} Comparison of the measured $R_{\zc}$ and
$R_{\zcp}$ with the theoretical predictions. 
}
 \begin{tabular}{c|c|c|c}
 \hline\hline
 Ratio        &Measurement         &        Tetraquark   &    Molecule   \\
 \hline
\multirow{8}{*}{$R_{\zc}$}  
&\multirow{8}{*}{$2.3\pm 0.8$ ~\cite{Collaboration:2017njt}} & $230_{-140}^{+330}$~\cite{Esposito:2014hsa} & $0.046_{-0.017}^{+0.025}$~\cite{Esposito:2014hsa} \\
&\multirow{8}{*}{} &  $0.27_{-0.17}^{+0.40}$~\cite{Esposito:2014hsa} & $1.78\pm0.41$~\cite{Goerke:2016hxf}  \\
&\multirow{8}{*}{}& 0.66~\cite{Faccini:2013lda} & $6.84\times10^{-3}$~\cite{Patel:2014zja}\\ 
&\multirow{8}{*}{}& $0.56\pm0.24$~\cite{Agaev:2016dev} & $0.12$~\cite{Ke:2013gia}\\ 
&\multirow{8}{*}{}& $0.95\pm0.40$~\cite{Dias:2013xfa}& \\
&\multirow{8}{*}{}& $1.08\pm0.88$~\cite{Wang:2017lot} &  \\
&\multirow{8}{*}{}&  $1.28\pm0.37$~\cite{Goerke:2016hxf}   &  \\
&\multirow{8}{*}{}&  $1.86\pm0.41$~\cite{Goerke:2016hxf}   &  \\
\hline
  $R_{\zcp}$&     $<1.2$~\cite{Ablikim:2013wzq}         
  &        $6.6_{-5.8}^{+56.8}$~\cite{Esposito:2014hsa}          & 
  $0.010_{-0.004}^{+0.006}$~\cite{Esposito:2014hsa} \\ 
 \hline\hline
 \end{tabular}
\end{table}

For $R_{\zcp}$, we can only report upper limits, but they are
smaller than the value calculated based on the tetraquark model. On the other hand, the upper limits are not
in contradiction with the molecule model calculation, which is
about two orders of magnitude smaller than the current upper
limits~\cite{Esposito:2014hsa}.

\acknowledgments

The BESIII collaboration thanks the staff of BEPCII and the IHEP computing center for their strong support. This work is supported in part by National Key Basic Research Program of China under Contract No. 2015CB856700; National Natural Science Foundation of China (NSFC) under Contracts Nos. 11335008, 11425524, 11625523, 11635010, 11735014, and 11575198; the Chinese Academy of Sciences (CAS) Large-Scale Scientific Facility Program; the CAS Center for Excellence in Particle Physics (CCEPP); Joint Large-Scale Scientific Facility Funds of the NSFC and CAS under Contracts Nos. U1532257, U1532258, U1732263; CAS Key Research Program of Frontier Sciences under Contracts Nos. QYZDJ-SSW-SLH003, QYZDJ-SSW-SLH040; 100 Talents Program of CAS; INPAC and Shanghai Key Laboratory for Particle Physics and Cosmology; German Research Foundation DFG under Contracts Nos. Collaborative Research Center CRC 1044, FOR 2359; Istituto Nazionale di Fisica Nucleare, Italy; Koninklijke Nederlandse Akademie van Wetenschappen (KNAW) under Contract No. 530-4CDP03; Ministry of Development of Turkey under Contract No. DPT2006K-120470; National Science and Technology fund; The Knut and Alice Wallenberg Foundation (Sweden) under Contract No. 2016.0157; The Swedish Research Council; U. S. Department of Energy under Contracts Nos. DE-FG02-05ER41374, DE-SC-0010118, DE-SC-0012069; University of Groningen (RuG) and the Helmholtzzentrum fuer Schwerionenforschung GmbH (GSI), Darmstadt.

\end{document}